\documentstyle[11pt,epsf]{article}
\def\DRAFT{0}
\renewcommand{\theequation}{\thesection.\arabic{equation}}
\if@twoside\oddsidemargin25mm
\else\oddsidemargin30mm\evensidemargin30mm\marginparwidth30mm\fi%
\begin{document}
\renewcommand{\theequation}{\thesection.\arabic{equation}}
%
%
\newcommand{\eq}{\begin{equation}}
\newcommand{\en}{\end{equation}}
\newcommand{\eqn}{$$}
\newcommand{\enn}{$$}
\newcommand{\eqa}{\begin{eqnarray}}
\newcommand{\ena}{\end{eqnarray}}
\newcommand{\eqan}{\begin{eqnarray*}}
\newcommand{\enan}{\end{eqnarray*}}
\newcommand{\bear}[1]{\begin{array}{#1}}
\newcommand{\enar}{\end{array}}
\newcommand{\spz}{\hspace{0.7cm}}
\newcommand{\lbl}[1]{ \ifnum \DRAFT = 0
                             {\label {#1}}
                      \else  {\makebox[0in]{\raisebox{-2ex}{\tiny #1}
                                            \hspace{-6ex}}}
                             {\label {#1}}
                      \fi}
\newcommand{\rf}[1]{  \ifnum \DRAFT = 0
                             \ref {#1}
                      \else {\ref {#1}}
                            \makebox[0in]{\raisebox{-2ex}{\tiny #1}}
                      \fi}
\newcommand{\ct}[1]{  \ifnum \DRAFT = 0
                             \cite {#1}
                      \else {\cite {#1}}
                            \makebox[0in]{\raisebox{-2ex}{\tiny #1}}
                      \fi}
\newlength{\superigor}
\newcommand{\spazio}[1]{\settowidth{\superigor}{${#1}$}
                        \makebox[\superigor]{} }
%
%
\def \eroinappendice{0}
\newcounter{temporaneo}
\newcounter{appendice}
\newcommand{\sect}[1]{
\ifnum \eroinappendice =1
       \def\eroinappendice{0}
       \setcounter{appendice}{\value{section}}
       \setcounter{section}{\value{temporaneo}}
       \renewcommand{\theequation}{\thesection.\arabic{equation}}
       \fi
\setcounter{equation}{0}\section{#1}}
\newcommand{\app}[1]{
\ifnum \eroinappendice =0
       \def\eroinappendice{1}
       \setcounter{temporaneo}{\value{section}}
       \setcounter{section}{\value{appendice}}
       \renewcommand{\thesection}{Appendix \Alph{section}: }
       \renewcommand{\theequation}{\Alph{section}.\arabic{equation}}
      \fi
\setcounter{equation}{0}\section{#1}}
\def\sqr#1#2{{\vcenter{\hrule height.#2pt
     \hbox{\vrule width.#2pt height#1pt \kern#1pt
        \vrule width.#2pt}
     \hrule height.#2pt}}}
\def\smallsquare{\mathchoice\sqr34\sqr34\sqr{2.1}3\sqr{1.5}3}
\def\square{\mathchoice\sqr68\sqr68\sqr{4.2}6\sqr{3.0}6}
\def\thinspace{\kern .16667em}
\def\Dir{\nabla\kern-2ex\Big{/}}
\def\dslash{\partial\kern-1.5ex\Big{/}}
\catcode64=11
\def\reali{{\hbox{\s@ l\kern-.5ex R}}}
\def\naturali{{\hbox{\s@ l\kern-.5ex N}}}
\def\interi{{\mathchoice
 {\hbox{Z\kern-1.5mm Z}}
 {\hbox{Z\kern-1.5mm Z}}
 {\hbox{{Z\kern-1.2mm Z}}}
 {\hbox{{Z\kern-1.2mm Z}}}  }}
\def\complessi{{\bf C}}
\def\unity{{\hbox{\s@ 1\kern-.8mm l}}}
\def\uno{{\hbox{ 1\kern-.8mm l}}}
\catcode64=12
\def\pd#1{{\partial~\over\partial #1}}
\def\part{\partial}
\def\rd{\sqrt{2}}
\def\um{{1\over2}}
\def\usrd{{1\over\sqrt{2}}}
\def\dxy{\delta(x-y)}
\def\dij{\delta^{ij}}
\def\Llrarr{\Longleftrightarrow}
\def\lorarr{\longrightarrow}
\def\lrarr{\leftrightarrow}
\def\rarr{\rightarrow}
\def\larr{\leftarrow}
\def\Rarr{\Rightarrow}
\def\Larr{\Leftarrow}
\def\ot{\otimes}
\def\ww{\wedge}
\def\dag{\dagger}
\def\inf{\infty}
\def\bk{\backslash}
\def\CA{{\cal A}}
\def\CB{{\cal B}}
\def\CC{{\cal C}}
\def\CF{{\cal F}}
\def\CG{{\cal G}}
\def\CI{{\cal I}}
\def\CL{{\cal L}}
\def\CM{{\cal M}}
\def\CN{{\cal N}}
\def\CP{{\cal P}}
\def\CS{{\cal S}}
\def\CT{{\cal T}}
\def\CU{{\cal U}}
\def\CV{{\cal V}}
\def\CY{{\cal Y}}
\def\CZ{{\cal Z}}
\def\aa{\alpha}
\def\bb{\beta}
\def\cc{\chi}
\def\cb{\bar\chi}
\def\dd{\delta}
\def\DD{\Delta}
\def\eb{\bar\epsilon}
\def\etab{\bar\eta}
\def\ff{\phi}
\def\fs{\phi^{\star}}
\def\fd{\phi^{\dagger}}
\def\FF{\Phi}
\def\vf{\varphi}
\def\gg{\gamma}
\def\GG{\Gamma}
\def\kk{\kappa}
\def\ll{\lambda}
\def\LL{\Lambda}
\def\lb{\bar\lambda}
\def\oo{\omega}
\def\OO{\Omega}
\def\Ob{\bar\Omega}
\def\OB{\bar\Omega}
\def\pp{\psi}
\def\PP{\Psi}
\def\pb{\bar\psi}
\def\PB{\bar\Psi}
\def\rr{\rho}
\def\rb{\bar\rho}
\def\ss{\sigma}
\def\SS{\Sigma}
\def\tt{\theta}
\def\tb{\bar\theta}
\def\taub{\bar\tau}
\def\xib{\bar\xi}
\def\zz{\zeta}
\def\zb{\bar\zeta}
\newcommand{\mat}[4]{\left(
                     \begin{array}{cc}
                     {#1} & {#2} \\
                     {#3} & {#4}
                     \end{array}
                     \right)
                    }
\newcommand{\vett}[2]{\left(
                      \begin{array}{c}
                     {#1} \\
                     {#2}
                     \end{array}
                     \right)
                    }
\newcommand{\vet}[2]{\left(
                     \begin{array}{cc}
                     {#1} &  {#2}
                     \end{array}
                     \right)
                    }
\newcommand{\ft}[3]{\int {d^{#1}{#2}\over (2\pi)^{#1}} ~ e^{i {#2}\cdot{#3}} }
\newcommand\modu[1]{|{#1}|}
%
\font\cmss=cmss10 \font\cmsss=cmss10 at 7pt
\def\twomat#1#2#3#4{\left(\matrix{#1 & #2 \cr #3 & #4}\right)}
\def\inbar{\vrule height1.5ex width.4pt depth0pt}
\def\IC{\relax\,\hbox{$\inbar\kern-.3em{\rm C}$}}
\def\ID{\relax{\rm I\kern-.18em D}}
\def\IL{\relax{\rm I\kern-.18em L}}
\def\IF{\relax{\rm I\kern-.18em F}}
\def\IH{\relax{\rm I\kern-.18em H}}
\def\II{\relax{\rm I\kern-.17em I}}
\def\IN{\relax{\rm I\kern-.18em N}}
\def\IP{\relax{\rm I\kern-.18em P}}
\def\IQ{\relax\,\hbox{$\inbar\kern-.3em{\rm Q}$}}
\def\bfzero{\relax\,\hbox{$\inbar\kern-.3em{\rm 0}$}}
\def\IR{\relax{\rm I\kern-.18em R}}
\def\ZZ{\relax\ifmmode\mathchoice
{\hbox{\cmss Z\kern-.4em Z}}{\hbox{\cmss Z\kern-.4em Z}}
{\lower.9pt\hbox{\cmsss Z\kern-.4em Z}}
{\lower1.2pt\hbox{\cmsss Z\kern-.4em Z}}\else{\cmss Z\kern-.4em
Z}\fi}
\def\bfone{\relax{\rm 1\kern-.35em 1}}
\def\dop{{\rm d}\hskip -1pt}
\def\real{{\rm Re}\hskip 1pt}
\def\trace{{\rm Tr}\hskip 1pt}
\def\ii{{\rm i}}
\def\diag{{\rm diag}}
\def\sch#1#2{\{#1;#2\}}
\def\bfone{\relax{\rm 1\kern-.35em 1}}
\font\cmss=cmss10 \font\cmsss=cmss10 at 7pt
\def\a{\alpha} \def\b{\beta} \def\d{\delta}
\def\e{\epsilon} \def\c{\gamma}
\def\G{\Gamma} \def\l{\lambda}
\def\L{\Lambda} \def\s{\sigma}
\def\cA{{\cal A}} \def\cB{{\cal B}}
\def\cC{{\cal C}} \def\cD{{\cal D}}
\def\cF{{\cal F}} \def\cG{{\cal G}}
\def\cH{{\cal H}} \def\cI{{\cal I}}
\def\cJ{{\cal J}} \def\cK{{\cal K}}
\def\cL{{\cal L}} \def\cM{{\cal M}}
\def\cN{{\cal N}} \def\cO{{\cal O}}
\def\cP{{\cal P}} \def\cQ{{\cal Q}}
\def\cR{{\cal R}} \def\cV{{\cal V}}\def\cW{{\cal W}}
%
%
%
\def\crr{\crcr\noalign{\vskip {8.3333pt}}}
\def\tilde{\widetilde}
\def\bar{\overline}
\def\us#1{\underline{#1}}
\let\shat=\hat
\def\hat{\widehat}
\def\hyp{\vrule height 2.3pt width 2.5pt depth -1.5pt}
\def\square{\mbox{.08}{.08}}
\def\Coeff#1#2{{#1\over #2}}
\def\Coe#1.#2.{{#1\over #2}}
\def\coeff#1#2{\relax{\textstyle {#1 \over #2}}\displaystyle}
\def\coe#1.#2.{\relax{\textstyle {#1 \over #2}}\displaystyle}
\def\half{{1 \over 2}}
\def\shalf{\relax{\textstyle {1 \over 2}}\displaystyle}
\def\dag#1{#1\!\!\!/\,\,\,}
\def\to{\rightarrow}
\def\notin{\hbox{{$\in$}\kern-.51em\hbox{/}}}
\def\shdot{\!\cdot\!}
\def\ket#1{\,\big|\,#1\,\big>\,}
\def\bra#1{\,\big<\,#1\,\big|\,}
\def\equaltop#1{\mathrel{\mathop=^{#1}}}
\def\Trbel#1{\mathop{{\rm Tr}}_{#1}}
\def\inserteq#1{\noalign{\vskip-.2truecm\hbox{#1\hfil}
\vskip-.2cm}}
\def\attac#1{\Bigl\vert
{\phantom{X}\atop{{\rm\scriptstyle #1}}\phantom{X}}}
\def\exx#1{e^{{\displaystyle #1}}}
\def\del{\partial}
\def\delbar{\bar\partial}
\def\nex#1{$N\!=\!#1$}
\def\dex#1{$d\!=\!#1$}
\def\cex#1{$c\!=\!#1$}
\def\eg{{\it e.g.}} \def\ie{{\it i.e.}}
\def\IE{\relax{{\rm I\kern-.18em E}}}
\def\cE{{\cal E}}
\def\rt{{\cR^{(3)}}}
\def\IGam{\relax{{\rm I}\kern-.18em \Gamma}}
\def\IGa{\IA}
\def\ii{{\rm i}}
\def\diag{{\rm diag}}
\def\hp{{\bf HP}^{4(m+3)}}
\def\omep{\omega^{\scriptscriptstyle +}}
\def\omepind#1{\omega^{\scriptscriptstyle +\hskip 2pt #1}}
\def\omem{\omega^{\scriptscriptstyle -}}
\def\omemind#1{\omega^{\scriptscriptstyle -\hskip 2pt #1}}
\def\omepm{\omega^{\scriptscriptstyle \pm}}
\def\omepmind#1{\omega^{\scriptscriptstyle \pm\hskip 2pt #1}}
\def\Omep{\Omega^{\scriptscriptstyle +}}
\def\Omepind#1{\Omega^{\scriptscriptstyle +\hskip 2pt #1}}
\def\Omem{\Omega^{\scriptscriptstyle -}}
\def\Omemind#1{\Omega^{\scriptscriptstyle -\hskip 2pt #1}}
\def\Omepm{\Omega^{\scriptscriptstyle \pm}}
\def\Omepmind#1{\Omega^{\scriptscriptstyle \pm\hskip 2pt #1}}
\def\Jp{{\cal J}^{\scriptscriptstyle +}}
\def\Jpind#1{{\cal J}^{\scriptscriptstyle +\hskip 2pt #1}}
\def\Jm{{\cal J}^{\scriptscriptstyle -}}
\def\Jmind#1{{\cal J}^{\scriptscriptstyle -\hskip 2pt #1}}
\def\Jpm{{\cal J}^{\scriptscriptstyle \pm}}
\def\Jpmind#1{{\cal J}^{\scriptscriptstyle \pm\hskip 2pt #1}}
\def\Jmp{{\cal J}^{\scriptscriptstyle \mp}}
\def\Jmpind#1{{\cal J}^{\scriptscriptstyle \mp\hskip 2pt #1}}
\def\Qpind#1{Q^{\scriptscriptstyle +\hskip 2pt #1}}
\def\Qmind#1{Q^{\scriptscriptstyle -\hskip 2pt #1}}
\def\Qpmind#1{Q^{\scriptscriptstyle -\hskip 2pt #1}}
\def\FFpind{F^{\scriptscriptstyle + }}
\def\FFmind{F^{\scriptscriptstyle -}}
\def\KKpind{K^{\scriptscriptstyle + }}
\def\KKmind{K^{\scriptscriptstyle -}}
\def\KKpmind{K^{\scriptscriptstyle \pm}}
\newtheorem{definizione}{Definition}
\newtheorem{domanda}{Question}
\newtheorem{risposta}{Answer}
\def\o#1#2{{#1}\over{#2}}
\newtheorem{affermazione}{Statement}
\begin{titlepage}
\hskip 5.5cm
\vbox{
\hbox{DFTT 31/96}
\hbox{IFUM 535/96 FT}
\hbox{NORDITA 43/96 P}
\hbox{SISSA 101/96/EP}
\hbox{hep-th/9607032}
\hbox{July, 1996}}
\vfill
\begin{center}
{\LARGE Spontaneous $N=2 \, \to \,  N=1$ \\
\vskip 1.5mm
local supersymmetry breaking \\
\vskip 1.5mm
with surviving compact gauge groups$^*$ }\\
\vfill
{\large Pietro Fr\'e $^1$, Luciano Girardello $^2$,
Igor Pesando$^1$
  and  Mario Trigiante$^3$   } \\
\vfill
{\small
$^1$ Dipartimento di Fisica Teorica, Universit\'a di Torino, via P. Giuria 1,
I-10125 Torino, \\
 Istituto Nazionale di Fisica Nucleare (INFN) - Sezione di Torino, Italy \\
and Nordita, Blegdamsvej 17, DK-1200 Copenhagen \ OE, Denmark\\
\vspace{6pt}
 $^2$ Dipartimento di Fisica, Universit\`a di Milano, via Celoria 6,
I-20133 Milano,\\
and Istituto Nazionale di Fisica Nucleare (INFN) - Sezione di Milano, Italy\\
\vspace{6pt}
$^3$ International School for Advanced Studies (ISAS), via Beirut 2-4,
I-34100 Trieste\\
and Istituto Nazionale di Fisica Nucleare (INFN) - Sezione di Trieste, Italy\\
 }
\end{center}
\vfill
\begin{center}
{\bf Abstract}
\end{center}
{\small Generic partial supersymmetry breaking of $N=2$ supergravity
with zero vacuum energy and
with surviving unbroken arbitrary gauge groups
is exhibited. Specific examples are given.}
\vspace{2mm} \vfill \hrule width 3.cm
{\footnotesize
 $^*$ Supported in part by   EEC  Science Program SC1*CT92-0789.}
\end{titlepage}
\setcounter{footnote}{0}
\section{Introduction}
$N=2$ supergravity and $N=2$ rigid gauge theory have recently played a major
role in the discussion of string--string dualities
\cite{stdua_1,stdua_2,stdua_3,stdua_4,stdua_5,stdua_6,stdua_7} and in the analysis
of the non--perturbative
phases of Yang--Mills theories \cite{SW_1,SW_2,SWmore_1,SWmore_2}. Furthermore in its
ten years long history, $N=2$ supergravity has attracted the interest of
theorists because of the beautiful and rich geometrical structure of its
scalar sector, based on the manifold:
\begin{equation}
{\cal M}_{scalar} = {\cal SK}_n \, \otimes \, {\cal Q}_m
\label{scalma}
\end{equation}
where ${\cal SK}_n$ denotes a complex $n$--dimensional special K\"ahler
manifold \cite{speckal_1,speckal_2,speckal_3,skgsugra_1,speckal_4} (for a review of
Special K\"ahler geometry see either \cite{pietrolectures} or
\cite{toinelectures}) and ${\cal Q}_m$
a quaternionic $m$--dimensional quaternionic
manifold, $n$ being the number of vector multiplets and $m$ the  number of
hypermultiplets \cite{quatgeom_1,quatgeom_2,quatgeom_3,skgsugra_1}.
\par
In spite of its beauty and of its significant role in understanding the
structure of non perturbative field and string theory, applications of
$N=2$ supergravity to the description of the real world have  been
hampered by the presence of mirror fermions
and by a tight structure which limits severely
the mechanisms of spontaneous breaking of local $N=2$ SUSY.
Any attempt to investigate the fermion problem requires a thorough
understanding of the spontaneous breaking with zero vacuum energy.
In particular, an interesting feature is the sequential breaking to $N=1$
and then to $N=0$ at two different scales. Sometimes ago
a negative result  on partial breaking was established within
the $N=2$ supergravity formulation based on conformal tensor calculus \cite{lucia_nogo}.
A particular way out was indicated in an ad hoc model \cite{newlucia} which prompted
some generalizations based on Noether couplings \cite{berianew}.
\par
 With the developments in special K\"ahler geometry
\cite{newspec_1,stdua_4,stdua_5} stimulated by the studies on
S--duality, the situation can now be cleared in general terms. Indeed
it has appeared from
\cite{n2break_1,n2break_2,n2break_3,n2break_4}  that the negative results
on $N=2$ partial supersymmetry breaking were the consequence of unnecessary
restrictions imposed on the formulation of special K\"ahler geometry and could
be removed.
\par
In this paper we discuss in detail the generic structure of partial breaking
within
the general form of $N=2$ supergravity  theory coupled to an
arbitrary number of vector multiplets and hypermultiplets and
with an arbitrary gauging of the scalar manifold isometries.
In particular an explicit example is worked out in section 3, based on
the choice for the vector multiplets of the special K\"ahler manifold
$SU(1,1)/U(1) \, \otimes \, SO(2,n)/SO(2) \times SO(n)$ and of the
quaternionic manifold $SO(4,m)/SO(4) \times SO(m)$ for the
hypermultiplets.\footnote{This choice of manifolds is inspired by
string theory since it corresponds to $N=2$ truncations of string
compactifications on $T^6$ or, more generally, to the moduli space
of various $(2,2)$ conformal field theories (for instance free
fermion constructions).}
\subsection{The bearing of the coordinate free definition of
Special K\"ahler geometry}
A main point in our subsequent discussion is the use of symplectic
covariance within the coordinate free definition of special K\"ahler
geometry \cite{speckal_3,skgsugra_1}.
\par
Adopting the conventions of \cite{fundpaper} a complex $n$--dimensional
Hodge--K\"ahler manifold is {\it special K\"ahler of the local type} if
there exists a bundle ${\cal H}={\cal SV}\otimes {\cal L}$ where
${\cal SV} \stackrel{\pi}{\longrightarrow} {\cal M}_n$ is a holomorphic
flat vector bundle of rank $2n+2$ with structural group $Sp(2n+2,\IR)$,
while ${\cal L} \stackrel{\pi}{\longrightarrow} {\cal M}_n$ is the line bundle
whose first Chern class equals the K\"ahler class ($K=c_1({\cal L}$) and
if there exists a symplectic section of ${\cal H}$, named $\Omega$
such that:
\begin{equation}
K = {\o {\rm i}{2 \pi}}\,  \partial \,{\bar \partial} \,
{\rm lg} \, \left ( {\rm i} \langle \Omega \vert {\bar \Omega } \rangle
\right ) \, \equiv \, {\o {\rm i}{2 \pi}}\,  \partial \,{\bar \partial} \,
{\rm lg} \, \left ( {\rm i} \Omega^T \, \left (
\matrix { {\bf 0}  & {\bf 1}  \cr
- {\bf 1}  & {\bf 0} \cr  }\right )
 {\bar \Omega}
\right )
\label{magic}
\end{equation}
the boldfaced blocks in the above equation being $(n+1)\times (n+1)$ matrices.
Naming $\{ z^{i} \} $ a set of coordinates for ${\cal M}_n$, the symplectic
section $\Omega$ is usually written as:
\begin{equation}
\Omega = \left ( \matrix { X^\Lambda (z) \cr F_\Sigma (z) } \right )
\label{specsec}
\end{equation}
where the capital Greek indices run on $n+1$ values and label the full set
of vector fields of the theory including the graviphoton ($\Lambda=0$).
The upper half of the symplectic section is therefore associated with the
electric gauge potentials and it could be defined as the coefficient of the
gravitino contribution to the supersymmetry variation of these fields:
\begin{equation}
\delta_{SUSY} A^\Lambda_\mu = \dots + {\rm const} \, e^{{\cal K}} \,
X^\Lambda (z) \, {\bar \epsilon}^A \, \psi_\mu^B \, \varepsilon_{AB}
\, + \dots
\label{susanna}
\end{equation}
where $e^{{\cal K}}$ denotes the exponential of the K\"ahler potential. The
lower part of the symplectic section $F_\Sigma (z)$, which enters
the construction of the supersymmetric lagrangian in several ways,
would play, in the supersymmetry transformation rule
of the dual magnetic potentials $A_{\Sigma\vert \mu}^{(magn.)}$, had
we introduced them, the same role as it is played by $X^\Lambda (z) $ in the
transformation rule of the  electric potentials.
\par
A built in feature of special geometry is the possibility of performing
symplectic transformations of $Sp(2n+2,\IR)$. Such transformations correspond
to changes of bases for the symplectic sections in the bundle ${\cal H}$. Under
such transformations the notion of electric and magnetic gauge fields changes
and in fact the $Sp(2n+2,\IR)$ are realized as {\it duality transformations}
which are never symmetries of the lagrangian but they can be symmetries of
the field equations plus Bianchi identities when they correspond to the
symplectic embedding of isometries of the scalar manifold. If the symplectic
transformations are not symmetries, they are anyhow conceivable field
redefinitions and the question is whether
they yield equivalent formulations of the same theory. The answer is that
such physical equivalence occurs only in the case of {\it ungauged theories},
namely when all the electric charges are set to zero and the gauge group is
abelian. Indeed it is only in this case that distinction between electric
and magnetic fields is immaterial. As soon as an electric current is introduced
the distinction is established and, at least at the classical (or semiclassical)
level, the only
symplectic transformations that yield equivalent theories
(or can be symmetries) are the perturbative ones generated by lower triangular
symplectic matrices:
\begin{equation}
\left ( \matrix{{\bf A} & {\bf 0} \cr {\bf C} & {\bf D}\cr }\right )\,
\label{perturb}
\end{equation}
\par
It follows that different bases of symplectic sections for the bundle
${\cal H}$ yield, after gauging, inequivalent physical theories. The
possibility of realizing or not realizing partial supersymmetry breaking
$N=2 \, \to \, N=1$ are   related to this choice of symplectic bases.
In the tensor calculus formulation the lower
part of the symplectic section $F_{\Lambda}(z)$ should be, necessarily,
derivable from a holomorphic prepotential $F(X)$ that is a degree two
homogeneous function of the upper half of the section:
\begin{equation}
F_{\Sigma}(z) = {\o {\partial }{\partial X^{\Sigma}(z)}} \, F(X(z))
\quad ; \quad
F\left ( \lambda \, X^{\Sigma}(z)\, \right ) = \,\lambda^2 \,
F\left (   X^{\Sigma}(z)\, \right )
\label{prepotent}
\end{equation}
This additional request is optional in the more general geometric
formulation of $N=2$ supergravity \cite{skgsugra_1,fundpaper} where only the intrinsic
definition of special geometry is utilized for the construction of the
lagrangian. It can be shown that the condition for the existence of the
holomorphic prepotential $F(X)$ is the non degeneracy of the
jacobian matrix $e^{I}_i(z) \equiv \partial_i\left (X^{I}/X^0\right );
I=1,\dots , n$. There are symplectic bases where this jacobian has vanishing
determinant and there no $F(X)$ can be found. If one insists on the
existence of the prepotential such bases are discarded {\it a priori}.
\par
There is however another criterion to select symplectic bases which has
a much more intrinsic meaning and should guide our choice. Given the
isometry group $G$ of the special K\"ahler manifold ${\cal SV}$, which is
an intrinsic information, which subgroup of $G_{class} \subset G$ is
realized by classical symplectic matrices
$\left ( \matrix{{\bf A} & {\bf 0} \cr {\bf 0} & {\bf D}\cr }\right )$
and which part of $G$ is realized by non perturbative symplectic matrices
$\left ( \matrix{{\bf A} & {\bf B} \cr {\bf C} & {\bf D}\cr }\right )$ with
${\bf B} \ne 0$ is a symplectic base dependent fact. It appears that in
certain cases, very relevant for the analysis of string inspired supergravity,
the maximization of $G_{class}$, required by a priori symmetry
considerations, is incompatible with the condition ${\rm det}
\left ( e^{I}_i(z) \right ) \, \ne \, 0$ and hence with the existence of
a prepotential $F(X)$.  If the unneccessary condition on $F(X)$ existence
is removed and the maximally symmetric symplectic bases are accepted
the no-go results on partial supersymmetry breaking can also be removed.
Indeed in \cite{n2break_1} it was shown that by gauging a group
\begin{equation}
G_{gauge}= \IR^2
\end{equation}
in a N=2 supergravity model with just one vector multiplet and one hypermultiplet
based on the scalar manifold:
\begin{equation}
{\cal SK}={\o{SU(1,1)}{U(1)}} \quad \quad ; \quad \quad {\cal Q}= {\o{SO(4,1)}{SO(4)}}
\label{giraferpor}
\end{equation}
supersymmetry can be spontaneously broken from $N=2$ down to $N=1$,
provided one uses the symplectic basis  where the embedding of
$SU(1,1)\equiv SL(2,\IR)$ in $Sp(4,\IR)$ is the following \footnote{ Here
$\eta$ is the standard constant metric with $(2,n)$ signature}:
\begin{equation}
 \forall \,   \left ( \matrix{a & b \cr  c &d \cr }\right )
\, \in \, SL(2,\IR) \quad {\stackrel{\iota_\delta}
{\hookrightarrow}}
\left ( \matrix{ a \, \bfone & b \, \eta \cr c \, \eta  &
d \, \bfone \cr } \right )
\, \in \, Sp(4,\IR)
\label{ortolettodue}
\end{equation}
\par
\subsection{Our results}
In the present paper we generalize the above result to the case of an
N=2 supergravity with $n+1$ vector multiplets and $m$ hypermultiplets
based on the scalar manifold:
\begin{equation}
{\cal SK}_{n+1} = {\o{SU(1,1)}{U(1)}} \otimes
{\o{SO(2,n)}{SO(2)\otimes SO(n)}} \quad \quad ; \quad\quad
{\cal Q}_m = {\o{SO(4,m)}{SO(4)\otimes SO(m)}}
\label{ourchoice}
\end{equation}
We show that we can obtain partial breaking $N=2 \, \to \, N=1$ by
\begin{itemize}
\item {choosing the Calabi--Vesentini symplectic basis where the embedding of
$SL(2,\IR)\otimes SO(2,n)$ into $Sp(4+2n,\IR)$ is the following:
\begin{eqnarray}
\forall \,   \left ( \matrix{a & b \cr  c &d \cr }\right )
\, \in \, SL(2,\IR) & {\stackrel{\iota_\delta}
{\hookrightarrow}} &
\left ( \matrix{ a \, \bfone & b \, \eta \cr c \, \eta  &
 d \, \bfone \cr } \right )
\, \in \, Sp(2n+4,\IR) \nonumber\\
\forall \,  L \, \in \, SO(2,n) & {\stackrel{\iota_\delta}
{\hookrightarrow}} &
\left ( \matrix{ L & {\bf 0}\cr {\bf 0} & (L^T)^{-1} \cr } \right )
\, \in \, Sp(2n+4,\IR)
\label{ortoletto}
\end{eqnarray}
namely where all transformations of the group $SO(2,n)$ are linearly
realized on electric fields}
\item {gauging a group:
\begin{equation}
G_{gauge}= \IR^2 \, \otimes \, G_{compact}
\end{equation}
where
\begin{eqnarray}
\IR^2 \, & \cap & SL(2,R)\otimes SO(2,n) = 0  \nonumber\\
\IR^2 & \subset & \mbox{ an abelian ideal of the solvable Lie
subalgebra } V \,  \subset  so(4,m) \nonumber\\
G_{compact} & \subset & SO(n) \subset   SO(2,n) \nonumber\\
G_{compact} & \subset & SO(m-1) \subset  SO(4,m)
\label{condizie}
\end{eqnarray}
namely where $\IR^2$ is a group of two abelian translations acting on the hypermultiplet
manifold but with respect to which the vector multiplets have {\it zero charge}, while
the compact gauge group $G_{compact}$ commutes with such translations
and has a linear action on both
the hypermultiplet and the vector multiplets.}
\end{itemize}
The solvable Lie subalgebra $V \subset so(4,m)$ mentioned in eq.~\ref{condizie} is
that subalgebra of the non--compact isometry group $G$ algebra such that,
according to Alekseevski \cite{aleks} (see also \cite{cecale,cecfergir,vanderseypen}),
the {\it quaternionic
homogeneous manifold} $G/H$ can also be identified with the group
manifold ${\rm exp} [ V]$.
\par
In the following two sections we derive the result summarized above
starting from the recently obtained complete form of N=2 supergravity
with general scalar manifold interactions \cite{fundpaper}.
We conclude the
present introduction with some physical arguments why the result
should be
obtained precisely in the way we have described.
\subsection{General features of the partial $N=2$ supersymmetry
breaking}
\par
To break supersymmetry from $N=2$ down  to $N=1$ we must break the
$O(2)$ symmetry that rotates one gravitino into the other. This
symmetry is
gauged by the graviphoton $A^0_\mu$. Hence the graviphoton must become massive.
At the same time, since we demand that $N=1$ supersymmetry should be
preserved,
the second gravitino must become the top state of an $N=1$ massive spin
$3/2$ multiplet which has the form $\left \{ \left ( {\o{3}{2}} \right ), {2 }
\left ( 1 \right ) , \left ( {\o{1}{2}} \right )  \right \}   $.
Consequently not only the graviphoton but also a second gauge boson
$A^1_{\mu}$
must become massive through ordinary Higgs mechanism. This explains
while the partial supersymmetry breaking involves the gauging of a
two--parameter group. That it should be a non compact $\IR^2$ acting as
a translation group on the quaternionic manifold is more difficult to
explain a priori, yet we can see why it is very natural.
  In order to obtain a Higgs mechanism for
the graviphoton $A_\mu^0$ and the second photon $A_\mu^1$ these vectors must
couple to the hypermultiplets.  Hence these two fields should
 gauge isometries of the quaternionic manifold ${\cal Q}$. This is obvious.
That such isometries should be translations is understood by observing that in this way
one introduces a flat direction in the scalar potential,
corresponding to the vacuum expectation value of the hypermultiplet scalar, coupled
to the vectors in such a way. Finally the need to use the correct
symplectic basis is explained by the following remark. Inspection of
the gravitino mass-matrix shows that it depends on both the momentum
map ${\cal P}^0_\Lambda (q)$ for the quaternionic action of the gauge group on
the hypermultiplet manifold and on the upper (electric) part of the
symplectic section $X^\Lambda(z)$. In order to obtain a
mass matrix with a zero eigenvalue we need a contribution from both $X^0$ and $X^1$ at the
breaking point, which can always be chosen at $z^i=0$, since the vector multiplet scalars
are neutral (this is a consequence of $\IR^2$ being abelian). Hence
in the correct symplectic basis we should have both $X^0(0) \ne 0$
and $X^1(0) \ne 0$. This is precisely what happens in the
Calabi--Vesentini basis for the special K\"ahler manifold $SU(1,1)/U(1) \otimes
SO(2,n) / SO(2) \otimes SO(n)$. Naming $y^{i}$ $(i=1,\dots , n)$ a standard set of complex
coordinates for the $SO(2,n)/ (SO(2)\otimes SO(n))$ coset manifold, characterized by linear
transformation properties under the $SO(2)\otimes SO(n)$ subgroup and naming $S$
the dilaton field, i.e. the complex coordinate spanning the coset manifold
$SU(1,1)/U(1)$, the explicit form of the symplectic section (\ref{specsec})
corresponding to the symplectic embedding (\ref{ortoletto}) is
(see \cite{newspec_1} and \cite{pietrolectures}):
\begin{equation}
\Omega \, = \, \left ( \matrix{ X^0 \cr X^1 \cr X^{i} \cr F_0 \cr F_1 \cr F_i\cr } \right ) \, = \,
\left ( \matrix{ {\o{1}{2} }(1+y^2)\cr  \cr {\rm i}{\o{1}{2} }(1-y^2)\cr \cr y^{i} \cr S \,
{\o{1}{2} }(1+y^2)\cr S \, {\rm i}{\o{1}{2} }(1-y^2)\cr
- \, S \, y^{i} \cr } \right )  \, {\stackrel{y\, \to\,  0}{\longrightarrow}}
\, \left ( \matrix{ {\o{1}{2} } \cr {\rm i}{\o{1}{2} } \cr 0 \cr {\o{1}{2} }\, S \cr
{\rm i}{\o{1}{2} } \, S \cr 0\cr } \right )
\label{calvesesec}
\end{equation}
\sect{Formulation of the  $N=2 \to N=1$ SUSY breaking problem}
As it is well understood in very general terms (see \cite{kilspinold}), a
classical vacuum of an $N=r$ supergravity theory preserving $p \le r$ supersymmetries
 in Minkowski  space--time  is just a constant scalar field
configuration $ \phi^I (x) = \phi^I_0 \, (I = 1,\dots ,\mbox{dim} \, {\cal M}_{scalar}) $
corresponding to an extremum of the scalar potential and such that it
admits $p$ {\it Killing spinors}. In this context, Killing spinors are covariantly constant
spinor parameters of the supersymmetry transformation
$\eta^A_{(a)}$ $(A=1,\dots,n)$, $(a=1,\dots,p)$ such that the
SUSY variation of the fermion fields in the bosonic background
$g_{\mu\nu}=\eta_{\mu\nu} \, , \, A^\Lambda_\mu=0\, , \,  \phi^I =
\phi^I_0$ is zero for each $\eta^A_{(a)}$:
\begin{eqnarray}
\delta_a \, \psi_{\mu A} & \equiv & {\rm i}\gamma_\mu \, S_{AB}(\phi_0)
\, \eta^B_{(a)} \, = \, 0\nonumber\\
\delta_a \xi^i & \equiv &\Sigma^i_A (\phi_0 )\, \eta^A_{(a)} \, = \, 0   \quad \quad (a=1,\dots,p)
\label{kilspigen}
\end{eqnarray}
In  eq.(~\ref{kilspigen})
  the spin ${\o{3}{2}}$ fermion shift  $S_{AB}(\phi)$ and
the spin ${\o{1}{2}}$ fermion shifts  $\Sigma^i_A (\phi)$ are the
non-derivative contributions to the supersymmetry transformation
rules of the gravitino $\psi_{\mu A}$ and of the spin one half fields $\xi^i$, respectively.
The integrability conditions of supersymmetry transformation rules
are just the field equations. So it actually happens that the
existence of Killing spinors, as defined by eq.(~\ref{kilspigen}), forces the
constant configuration $\phi^I = \phi^I_0$ to be an extremum of the scalar potential.
\par
Hence we can just concentrate on the problem of solving eq.(~\ref{kilspigen}) in the
case $N=2$ with $p=1$.
\par
In $N=2$ supergravity there are two kinds of spin one half fields:  the gauginos
$\lambda^{j^\star}_A$ carrying an $SU(2)$ index $A=1,2$ and a world--index
$j^\star=1^\star,\dots , n^\star$
of the tangent bundle $T^{(0,1)}{\cal SK}$ to the special K\"ahler manifold
($n=\mbox{dim}_{\bf C}\, {\cal SK}=
\# \mbox{vector multiplets}$ ) and the
hyperinos $\zeta^\alpha$ carrying an index $\alpha$=$(1,\dots,n)$ running in the
fundamental representation of $Sp(2m,\IR)$  ($m=\mbox{dim}_{\bf Q}\, {\cal Q}=
\# \mbox{hypermultiplets}$ ). Hence there are three kind of fermion
shifts:
\eqa
\dd\psi_{\mu A}&=& i\gg_{\mu} S_{A B} \eta^{B}
\nonumber\\
\dd(g_{i^* j} \ll^{i^*}_{A}) &=& W_{j |A B} \eta^{B}\nonumber \\
\dd \zz^{\aa} &=& N_A^{\aa} \eta^{A}
\label{ciliegina}
\ena
According to the analysis and the conventions of \cite{fundpaper,skgsugra_1},
the shifts are
expressed in terms of the fundamental geometric structures defined over the special
K\"ahler and quaternionic manifolds as follows:
\begin{eqnarray}
S_{AB}&=&-\um {\rm i} (\ss_{x}\epsilon )_{A B} \CP^{x}_{\LL} L^{\LL}\nonumber \\
W_{j|A B}&=&- ({\epsilon_{A B}} \part_{j}\CP_{\LL} L^{\LL} +{\rm i}
(\ss_{x}\epsilon)_{A B}
 \CP^{x}_{\LL} f^{\LL}_{j})\nonumber\\
 N_{A}^{\aa}& =& -2 {\bf i}_{\vec k_{\LL}}\CU^{\aa}_{A}  L^{\LL} \, =
 \, -2 \, {\cal U}^\alpha_{A \vert u} \, k^u_{\Lambda} \, L^\Lambda
\label{shifmatrici}
\end{eqnarray}
In eq.(\ref{shifmatrici}) the index $u$ runs on $4m$ values
corresponding to any set of of $4m$ real coordinates
for the quaternionic manifold. Further $\CP_{\LL}$ is the holomorphic momentum map
for the action of the gauge group ${\cal G}_{gauge}$ on the special
K\"ahler manifold ${\cal SK}_n$, $\CP^{x}_{\LL} \, (x=1,2,3)$ is the
triholomorphic momentum map for the action of the same group on the
quaternionic manifold, $L^{\Lambda}= e^{{\cal K}/2} X^\Lambda(z)$ is
the upper part of the symplectic section (\ref{specsec}), rescaled
with the exponential of one half the K\"ahler potential ${\cal K}(z,{\bar z})$
(for more
details on special geometry see \cite{fundpaper},
\cite{pietrolectures}, \cite{toinelectures}), ${\vec k}_{\LL}$ is the
Killing vector generating the action of the gauge group on both
scalar manifolds and finally $\CU^{\aa}_{A}$ is the vielbein 1--form on
the quaternionic manifold carrying an $SU(2)$ doublet index and an
index $\alpha$ running in the fundamental representation of
$Sp(2m,\IR)$.
\par In the case of effective supergravity theories
that already take into account the perturbative and non--perturbative
quantum corrections of string theory the manifolds ${\cal SK}$ and
${\cal Q}$ can be complicated non--homogeneous spaces
without continuous isometries. It is not in
such theories, however, that one performs the gauging of non abelian
groups and that looks for a classical breaking of supersymmetry.
Indeed, in order to gauge a non abelian group, the scalar manifold
must admit that group as a group of isometries. Hence ${\cal
M}_{scalar}$ is rather given by the homogeneous coset manifolds that
emerge in the {\it field theoretical limit} to {\it the tree level
approximation} of superstring theory. In a large variety of models the
tree level approximation yields the choice (~\ref{ourchoice}) and we
concentrate on such a case to show that a constant configuration with
a single killing spinor can be found. Yet, as it will appear
from our subsequent discussion, the key point of our construction
resides in the existence of an $\IR^2$ translation isometry group on
${\cal Q}$ that can be gauged by two vectors associated with section
components $X^0 \, X^1$ that become  constants in the vacuum
configuration of ${\cal SK}$. That these requirements can be met
is a consequence of the algebraic structure \'a la Alekseevski of both
the special K\"ahlerian and the quaternionic manifold. Since such
algebraic structures exist for all homogeneous special and
quaternionic manifolds, we are lead to conjecture that the partial
supersymmetry breaking described below can be extended to most N=2
supergravity theories on homogenous scalar manifolds.  \par For the
rest of the paper, however, we concentrate on the study of case
(\ref{ourchoice}).
\par In the Calabi-Vesentini basis
(\ref{calvesesec}) the origin $y=0$ of the vector multiplet manifold
$SO(2,n)/SO(2) \otimes SO(n)$ is a convenient point where to look for
a configuration breaking $N=2 \, \to \, N=1$.  We shall argue that for
$y=0$ and for an arbitrary point in the quaternionic manifold $\forall
q \, \in \, SO(4,m)/SO(4) \times SO(m)$ there is always a suitable
group $\IR^2_q$ whose gauging achieves the partial supersymmetry
breaking. Actually the group $\IR^2_q$ is just the conjugate, via an
element of the isometry group $SO(4,m)$, of the group $\IR^2_0$ the
achieves the breaking in the origin $q=0$. Hence we can reduce the
whole analysis to a study of the neighborhood of the origin in both
scalar manifolds.  To show these facts we need to cast a closer look
at the structure of the hypermultiplet manifold.
\sect{Explicit solution}
\subsection{The quaternionic manifold $SO(4,m) / SO(4) \ot SO(m)$}
We start with the usual parametrization of the coset
$SO(4,m)\over SO(4)\otimes SO(m)$ \cite{cosetto}:
\eq
\IL(q)=
\left(
\bear{cc}
\sqrt{1+qq^t} & q \\
q^t &\sqrt{1+q^tq}
\enar
\right)
=\mat{r_1}{q}{q^t}{r_2}
\label{parametrizzazione}
\en
where $q=||q_{a t}||$ is a $4\times m$ matrix
\footnote{In the following letters from the beginning of the alphabet
will range over $1\dots 4$ while letters  from the end of the alphabet
will range over $1\dots m$}.
This coset manifold has a riemannian structure defined by the
vielbein, connection and  metric given below \footnote{For more details
on the following formulae see Appendix C.1 of ref.~ \cite{fundpaper}}:
\eqa
\IL^{-1} d\IL &=&
\left(
\bear{cc}
\tt & E \\
E^t  & \DD
\enar
\right) \in so(4,m) \nonumber \\
ds^2 &=& E^t \otimes E
\ena
The explicit form of the vielbein and connections which we utilise in the sequel
is:
\begin{equation}
E \, = \, r_1 dq -q dr_2 \quad ; \quad
\tt \, = \, r_1 dr_1 -q dq^t \quad ; \quad
\DD \, = \, r_2 dr_2 -q^t dq
\end{equation}
where $E$ is the coset vielbein, $\theta$ is the $so(4)$-connection
and $\Delta$ is the $so(n)$-connection.
The quaternionic structure of the manifold is given by
\begin{equation}
K^x \, = \,  \um tr(E^t \ww J^x E) \quad ; \quad
\oo^x \, = \,-\um tr (\tt J^x) \quad ; \quad
\CU^{A \aa} \, = \, \usrd E^{a t} (e_a)^A_B
\lbl{3.4}
\end{equation}
$K^x$ being the triplet of hyperK\"ahler 2--forms, $\omega^x$ the
triplet of $su(2)$--connection 1--forms and ${\cal U}^{A\alpha}$ the
vielbein 1--form in the symplectic notation. Furthermore
$J^x$ is the triplet of $4\times 4$  self--dual 't Hooft matrices
$J^{+|x}$ normalized as in \cite{fundpaper}, $e_a$ are the quaternionic units
as given in \cite{fundpaper} and   the symplectic index $\alpha=1,\dots , 2m$
is identified with a pair of an $SU(2)$ doublet index $B=1,2$ times an $SO(4)$ vector
index $t=1,\dots, m$: $\aa\equiv B t$.
Notice also the factor $\um$ in the definition of $K^x$ with respect to the
conventions used in \cite{fundpaper},  which is necessary in order to have
$ \nabla \oo^x= -K^x$
\subsection{Explicit action of the isometries on the coordinates and the
killing vectors.}
Next we compute the killing vectors; to this purpose we need to
know the action of an element $g\in so(4,m)$ on the coordinates $q$.
To this effect we make use of the standard formula
\eq
\dd \IL \equiv k^{at}_{g} \, {\o{\partial}{\partial q^{at}}}= g \IL-\IL w_g
\label{cosettario}
\en
where
\eq
w_g=
\left(
\bear{cc}
w_1 &  \\
 & w_2
\enar
\right) \in so(4) \oplus so(m)
\en
is the right compensator and the element $g\in so(4,m)$ is given by
\eq
g=
\left(
\bear{cc}
a & b \\
b^t & c
\enar
\right)\in {\bf so(4,m)}
\label{donpeppone}
\en
with $a^t=-a$, $c^t=-c$.
The solution to (\ref{cosettario}) is:
\eqa
\dd q =  a q + b r_2 - q c -q \hat w_2 & = & a q +r_1 b -q c + \hat w_1 q\nonumber\\
\dd r_1  = [ a, r_1 ] + b q^t - r_1  \hat w_1 & ; &
\dd r_2  = [ c, r_2 ] + b^t q - r_2  \hat w_2\nonumber\\
w_1  =  a + \hat w_1 & ; &
w_2  =  c+\hat w_2
\label{cosevarie}
\ena
The only information we need to know about $\hat w_1, \hat w_2$ is that
they depend linearly on $b, b^t$.
Anyhow for completess we give their explicit form:
\eqa
\hat w_{1; a b}
&=& \left. { d (\sqrt{ \uno + x})_{a b}\over d x_{c d}} \right|_{x=qq^t}
(bq^t-qb^t)_{c d}\nonumber\\
\hat w_{2; s t}
&=& \left. { d (\sqrt{ \uno + y})_{s t}\over d y_{p q}} \right|_{y=q^tq}
(b^tq-q^tb)_{p q}
\ena
>From these expressions  we can obtain the Killing vector field
\eq
\vec k_g =  (\dd q)_{a t} \pd {q _{a t}}
\label{assassino}
\en
\subsection{The momentum map.}
We are now in a position to compute the triholomorphic momentum map ${\cal P}^x_g$
associated with the generic element \ref{donpeppone} of the $so(4,m)$ Lie algebra.
Given the vector field \ref{assassino}, we are supposed to solve the first order
linear differential equation:
\eq
{\bf i}_{{\vec k}_g} K^x = -\nabla \CP ^x
\en
$\nabla$ denoting the exterior derivative covariant with respect to
the $su(2)$--connection $\omega^x$ and ${\bf i}_{{\vec k}_g} K^x$ being the contraction
of the 2--form $K^x$ along the Killing vector field ${\vec k}_g$:
By direct verification the general solution is given by:
\begin{equation}
\CP^x_g =\um tr(\mat{J^x}{0}{0}{0} C_{g} )= tr(J^x P_{g})
\label{fragolina}
\end{equation}
where for any element  of the ${\bf so(4,m)}$ Lie algebra conjugated
with the adjoint action of the coset representative (\ref{parametrizzazione}), we have
introduced the following block decomposition and notation:
\eq
\forall { g}\, \in \, { so(4,m)}: \quad C_{ g}\,  \equiv \, \IL(q)^{-1} g
\IL(q) = \mat{2P_{ g}}
{ {\bf i}_{\vec k}E_{ g} }{ {\bf i}_{\vec k}E^t_{ g} }{ 2 Q_{ g}}
\label{fondamentale}
\en
Furthermore if we decompose $g=g^\Lambda \, {\bf T}_\Lambda$
the generic element $g$ along a basis  $\{ {\bf T}_\Lambda \}$ of
generators of the $so(4,m)$ Lie algebra, we can write:
\eq
\CP^x_\Lambda =\um tr(\mat{J^x}{0}{0}{0} C_{{\bf T}_\LL})= tr(J^x P_{{\bf T}_\LL})
\label{fragolona}
\en
\subsection{Solution of the breaking problem in a generic point of the
quaternionic manifold}
At this stage we can attempt to find a solution for our problem i.e. introducing
a gauging that yields a partial supersymmetry breaking.
\par
Recalling the supersymmetry variations of the Fermi fields (\ref{ciliegina})
  we evaluate them at the origin of the special K\"ahler
using the Calabi-Vesentini coordinates (\ref{calvesesec}):
\eqa
S_{A B}|_{y=0} &=& -{1\over 4} i (\ss_x \epsilon)_{A B} tr(J^x(P_0+i P_1))\nonumber\\
W_{ s| AB }|_{y=0} &=& -{1\over 4 (Im S)^{3\over 2}} i (\ss_x\epsilon)_{A B}
tr(J^x(P_0+i P_1))\nonumber\\
W_{ \aa | A B}|_{y=0} &=& 0\nonumber \\
N_A^\aa |_{y=0}&\propto& ({\bf i}_{0}E^{a t} -i {\bf i}_{1}E^{a t}_1) (e_a)^A_B
\label{variations}
\ena
where in the last equation we have identified $\aa\equiv a B$ as
explained after eq. (\rf{3.4}).
\par
In the next subsection we  explicitly compute a solution for the matrices $P_0,
P_1$ and ${\bf i}_{0}E^{a t}, {\bf i}_{1}E^{a t}$ at the origin of the
quaternionic manifold $\cQ_{m}$.
Starting from this result it can be shown that any point $q\neq 0$ can define
a vacuum of the theory in which SUSY is broken to $N=1$, provided a suitable
gauging is performed. Indeed we can find the general solution for any point $q$ by requiring
\eq
C_\LL(q)=C_\LL(q=0) ~~~~ \LL=0,1
\en
that is the group generators of the group we are gauging at a generic
point $q$  are given by
\eq
T_{\LL}(q)= \IL(q) T_\LL(0) \IL^{-1}(q)
\en
This result is very natural and just reflects the homogeneity of the
coset manifold,
i.e. all of its points are equivalent.
\subsection{Solution of the problem at the origin of $\cQ_{m}$.}
In what follows all the earlier defined quantities, related to
$\cQ_{m}$, will be computed near the origin $q=0$. The right-hand side
of equations (\ref{cosevarie},\ref{assassino}) is expanded in powers
of $q$ as it follows:
\eqa \dd q = b+aq-qc +O(q^2) & ; & \vec k_a \, =
\, (aq)_{a t} \pd {q _{a t}}\nonumber\\ \vec k_b \, =\, b_{a t} \pd
{q_{a t}} & ; & \vec k_c \, =\, -(qc)_{a t} \pd {q _{a t}}
\label{umamma}
\ena
The expressions for the vielbein, the connections and the quaternionic
structure, to the approximation order  we work, are:
\eqa
E \, =\,  dq-\um q dq^t q &;&
\tt\, =\,  \um dq q^t -\um q dq^t \nonumber\\
K^x &=&  tr\left( dq^t ~J^x~ dq - dq q^t ~ J^x ~ dq q^t \right) \nonumber\\
\oo^x \, =\,  -\um tr \left(dq q^t J^x \right) & ; &
\CU^{A \aa} \, =\,  \usrd dq^{a t} (e_a)^A_B
\lbl{partenza}
\ena
Finally the triholomorphic momentum maps corresponding to the a,b,c generators, have
the following form,  respectively :
\begin{equation}
\CP^x_a \, =\,  \um tr( J^x a + J^x q t^t a) \quad ; \quad
\CP^x_b \, =\, -\um tr (J^x q b^t) \quad ; \quad
\CP^x_c \, =\, -\um tr (J^x q c q^t)
\label{tripappa}
\end{equation}
Inserting eq.s (\ref{tripappa}) into equations (\ref{variations}) one finds the expressions
for the shift matrices  in the origin:
\eqa
S_{A B}|_{y=0} &=& -{1\over 4} i (\ss_x\epsilon)_{A B} tr(J^x(a_0+i a_1))\nonumber\\
W_{ s| AB }|_{y=0} &=& -{1\over 4 (Im S)^{3\over 2}} i (\ss_x\epsilon)_{A B}
tr(J^x(a_0+i a_1)) \nonumber\\
W_{ \aa | A B}|_{y=0} &=& 0 \nonumber\\
N_A^\aa |_{y=0} &\propto& (b_0^{a t} -i b^{a t}_1) (e_a)^A_B
\lbl{variatorig}
\ena
$a_{0,1}$ and $b_{0,1}$ being the a and b blocks of the matrices $P_{0,1}$,
respectively.
\par
It is now clear that in order to
break  supersymmetry we need $a\neq 0$ and $b \neq 0$ .
 From the first  two equations and from the requirement that the
gravitino mass--matrix $S_{AB}$ should have a zero
eigenvalue we get
\eq
\sum_x (a_0+i a_1)^2_x=0
\Rarr \vec a_0 \cdot \vec a_1= (\vec a_0)^2-(\vec a_1)^2=0
\en
We solve this constraint by setting ($ a_x =-{1\over 4} tr(J^x a)$ )
\eq
a_{0 x}= g_0 \dd_{x 1} ~~~~
a_{1 x}= g_1 \dd_{x 2}
\label{doncamillo}
\en
with $g_0=g_1$. These number are the gauge coupling constants of the
repeatedly mentioned gauge group $\IR^2$.
Note that due to the orthogonality of the antiself--dual t'Hooft matrices
${\bar J}^x$ to the self--dual ones $J^x$, the general solution is
not as in eq.~\ref{doncamillo}, but it involves additional arbitrary
combinations of  ${\bar J}^x$, namely:
\begin{equation}
a_0 \, =\,  g_0 J^1 + \bar a_{0 x} \bar {J}^x  \quad ; \quad
a_1 \, =\,  g_1 J^2 + \bar a_{1 x} \bar {J}^x
\end{equation}
For the conventions see \cite{fundpaper}.
To solve the last  of eq.s (\rf{variations}) we set
\eq
b_0 =\left(\bear{c} 0 \\ \vec \bb_0 \\ 0 \\ 0 \enar \right)
 ~~~~
b_1 =\left(\bear{c} 0 \\ 0 \\ \vec \bb_1 \\ 0 \enar \right)
\en
where $\vec \bb_0$ and $\vec \bb_1$ are $m$--vectors of the
in the fundamental representation of $SO(m)$.
\par
Now the essential questions are:
\begin{enumerate}
\item{\it  Can one find two commuting matrices belonging to
the ${\bf so(4,m)}$ Lie algebra and   satisfying the previous
constraints? These matrices are the generators of $\IR^2$.}
\item{\it If so, which is the maximal compact subalgebra ${\bf G}_{compact}\subset {\bf so(4,m)}$
commuting with them, namely the maximal compact subalgebra of the centralizer $Z(\IR^2)$?
${\bf G}_{compact}$ is the Lie algebra of the maximal compact gauge group that can survive
unbroken after the partial supersymmetry breaking. }
\end{enumerate}
\par
Let $a_{2},b_2,c_2$ denote the blocks of the commutator $[g_0,g_1]$.
To answer the first question we begin to seek a solution with $c=0$,
and it is easily checked that also their commutator have vanishing block $c_3$.
Let us now look at the condition $a_2=0$, namely
\begin{equation}
a_2= [a_0, a_1]+ b_0 b_1^t -b_1 b_0^t =
2 g_0 g_1 J^3 +2 \epsilon^{x y z} \bar a_0^x \bar a_1^y \bar J^z
+\um \vec \bb_0 \cdot \vec \bb_1 (J^3 + \bar J^3) = 0
\end{equation}
This equation can be solved if we set
\begin{equation}
\bar a_0^x \, =\,  \gg g_0 \dd_{x 1} \quad ; \quad
\bar a_1^x \, =\,  \gg g_1 \dd_{x 2} \quad ; \quad
\vec \bb_0 \cdot \vec \bb_1 =-4 g_0 g_1
\lbl{eq1}
\end{equation}
and if $\gamma^2=1 \, \leftarrow \, \gamma=\pm 1$.
Now we are left to solve the equation $b_2=a_0 b_1 -a_1 b_0= 0$, which
implies
\eq
(\gamma-1)g_0\vec{b}_{1}=(\gamma-1)g_1\vec{b}_{0}
\lbl{eq2}
\en
This equation is automatically fulfilled if $\gamma=1$ for any choice of
the $\vec{b}_{i}$s, while it is not consistent with the last of
equations (\ref{eq1}) if $\gamma = - 1$.
  Thus the only admissible value for $\gamma$ is $1$.
\par
Choosing $\vec \bb_0= (\bb_0,0 \dots 0)$ we can immediately answer the
second question: for the normalizer of the $\IR^2$ algebra we have
 $Z_{\bf so(4,m)}\left (\IR^2 \right )= {\bf so(m-1)}$, so that of the $m$
hypermultiplets one is eaten by the superHiggs mechanism and the remaining
$m-1$ can be assigned to any linear representation of a compact gauge group
that can be as large as $SO(m-1)$.
\section{The $N=2 \to N=1$ breaking problem in
Alekseevski\v{i}'s formalism.}
There are two main motivations for the choice of Alekseevski\v{i}'s
formalism \cite{aleks},\cite{cecale}
while dealing with the partial breaking of $N=2$ supersymmetry:
\begin{itemize}
\item {it provides a description of the quaternionic manifold as a group
manifold on which the generators of the $\IR^{2}$ act as traslation
operators on two (say $t_{0},t_{1}$) of the $4m$ scalar fields,
 parametrizing $\cQ_{m}$;}
\item {as it will be apparent in the sequel, the fields  $t_{0},t_{1}$ define flat
directions for the scalar potential. They
can be identified with the {\it hidden sector}
of the theory, for their coupling to the gauge fields $A_\mu^0$ and $A_\mu^1$ is
what determines the partial supersymmetry breaking.}
\end{itemize}
Thus Alekseevski\v{i}'s description of quaternionic manifolds provides a
conceptually very powerful tool to deal  with the SUSY
breaking problem, even in
the case in which $\cQ$ is not a symmetric homogeneous manifold \cite{beria}.
In what follows a sketchy overview of Alekseevski\v{i}'s formalism applied to
 the symmetric quaternionic manifold $\cQ_{m}$ will be given, and in terms of the corresponding
 quaternionic algebra, an explicit realization of the generators of the gauge group will be found.
\subsection{The Quaternionic Algebra and Partial SUSY Breaking.}
The starting point of Alekseevski\v{i}'s description of quaternionic manifolds
is a theorem stating that every Riemannian manifold $\cM$ with a transitive isometry group $\cG$
generated by a solvable Lie Algebra G (named a {\it normal} manifold), can be
 expressed as a group manifold
whose Lie Algebra $G^\prime $ is a suitable subalgebra of $G$: $\cM=e^{G^\prime}$
Therefore there exists a one-to-one correspondence between normal Riemmanian
manifolds and solvable Lie Algebras (endowed with an euclidean metric).
The quaternionic manifolds $\cQ_{m}$ we are dealing with  are normal and
therefore they can be obtained exponentiating a certain solvable subalgebra
 $V\subset {\bf so(4,m)}$, called {\it quaternionic algebra}. Alekseevski\v{i}
 classified the quaternionic manifolds in terms of the corresponding quaternionic
algebras.
 Every quaternionic algebra $V$ is characterized by a {\it quaternionic structure}
 which is an algebra of antisymmetric endomorphisms of $V$ generated by three
 operators $J_{1},J_{2},J_{3}$ fulfilling the standard algebra of quaternionic
imaginary units: $J_{\alpha} \dot J_{\beta}=- \delta_{\alpha\beta} \, {\rm Id}
+ \varepsilon_{\alpha\beta\gamma}J_{\gamma}$.
  Moreover a single antisymmetric
 endomorphism $J$ of a vector space such that $J^{2}=-Id$ is usually called a {\it
 complex structure}. A {\it K\"ahlerian Algebra} is defined as the solvable Lie algebra
 generating a normal K\"ahlerian manifold, and it is characterized by a complex structure $J$.
A common feature of the quaternionic algebras $V_{m}$ generating $\cQ_{m}$ is that
they have the form $V_{m}=U\oplus\tilde{U}$ where:
\eqa
U \, = \, F_{0}\oplus K_{m} & ; &
K_{m}\, = \, F_{1}\oplus F_{2}\oplus F_{3}\oplus Z\nonumber\\
\tilde{U}\, = \, J_{2}U & ; &
J_{1}U \, = \, U
\label{strutto}
\ena
The subspaces $F_{i}\quad i=0,1,2,3$, endowed with the complex structure $J_{1}$ are
2-dimensional K\"ahlerian algebras generated respectively by $\{h_{i},
g_{i}\}, \quad [h_{i},g_{i}]=g_{i} \quad i=0,1,2,3$. The space $Z$ and its image
$\tilde{Z}$ through $J_{2}$ in $\tilde{U}$ are the only parts of the whole algebra whose
dimension depends on $m= \# ~of~ hypermultiplets$, and thus it is natural to choose
the fields parametrizing them in some representation of $\cG_{compact}$, for, as it was
shown earlier, $\cG_{compact}\subset SO(m-1)$. At  fixed $m$ the $Z$-sector
can provide enough scalar fields as to fill a representation of the compact
gauge group. On the other hand the fields of the {\it hidden sector}
are to be chosen in the orthogonal complement with respect to $Z$ and $\tilde{Z}$  and
to be singlets with respect to  $\cG_{compact}$. Indeed, by
definition the fields of the hidden sectorinteract only with $A_\mu^0$ and $A_\mu^1$ \\
 The relations characterizing a generic quaternionic algebra  $V=U\oplus\tilde{U}$ are:
\begin{equation}
[U,U]\subset U ;\hspace{1.5 cm}[U,\tilde{U}]\subset \tilde{U};\hspace{1.5
cm}[\tilde{U},\tilde{U}]\subset \{ g_{0}\}
\label{relazioni}
\end{equation}
Using Alekseevski\v{i}'s notation, an orthonormal basis for U is provided
by $\{h_{i},g_{i}\quad (i=0,1,2,3), z^{\pm}_{k} (k=1,...,m-4)\}$, while for
$\tilde{U}$ an orthonormal basis is given by
 $\{p_{i},q_{i}\quad (i=0,1,2,3), \tilde{z}^{\pm}_{k}\quad (k=1,...,m-4)\}$.
All these generators are uniquely defined by the algebraic structure of
$V_{m}$ which was determined in Alekseevski\v{i}'s  work \cite{aleks}, and have a simple
representation in terms of the canonical basis of the full isometry algebra ${\bf so(4,m)}$.
\par
The next step is to determine  within this formalism, the generators of the
gauge group. As far as the $\IR^{2}$ factor is concerned we demand it to act
by means of translations on the coordinates of the coset representative $\IL(q)$. To attain this
purpose
the generators of the translations $T_{0},T_{1}$ will be chosen within
an abelian ideal $\cA\subset V_{m}$ and the representative of $\cQ_{m}$ will
be defined in the following way:
\eqa
\IL(t,b)\, =\, e^{T(t)}e^{G(b)} \quad (t) & = &t^{0},t^{1} \quad ;
\quad (b)=b^{1},...,b^{4m-2}\nonumber\\
T(t)\, =\,t^{\Lambda}T_{\Lambda}\quad \Lambda=0,1& ; &
G(t)\, =\, b^{a}G_{i}\quad a=1,...,4m-2\nonumber\\
V_{m}&=&T\oplus G
\label{margarina}
\ena
It is apparent from eq.(\ref{margarina}) that the left action of a transformation generated by $T_{\Lambda}$ amounts to a translation of the coordinates $t^{0},t^{1}$.
The latter  will define flat directions for the scalar potential.
As the scalar potential depends on the quaternionic
coordinates only through $\cP^{x}_{\alpha}$ and the corresponding killing vector, to prove
the truth of the above statement
it suffices to show that the momentum-map $\cP^{x}_{\alpha}(t,b)$ on $\cQ_{m}$
does not depend on the variables $t^{\Lambda}$.
Indeed a general expression for $\cP^{x}_{\alpha}$ associated to the generator
$T_{\alpha}i$ of ${\bf so(4,m)}$ is given by equation (\ref{fragolina}):
\eq
\cP^{x}_{\alpha}(t,b)=\frac{1}{2}tr(\IL^{-1}(t,b)T_{\alpha}\IL(t,b) {\cal J}^{x})=
\frac{1}{2}tr(e^{-G}e^{-T}T_{\alpha}e^{T}e^{G}{\cal J}^{x})
\label{masterfrutta}
\en
If $T_{\alpha}$ is a generator of the gauge group, then either it is in $T$
or it is in the compact subalgebra. In both cases it commutes with $T$
 allowing the exponentials of $T(t)$ to cancel against each other.
It is also straightforward to prove that the Killing vector components
on $\cQ_{m}$ do not depend on $t^{\Lambda}$.
\par
A maximal abelian ideal $\cA$ in $V_{m}$ can be shown to have dimension $m+2$.
Choosing
\begin{equation}
{\cal A}=\left\{ e_{1}, g_{3}, p_{0}, p_{3},q_{1},q_{2},\tilde{z}_{+}^{k} \right \} \quad
\quad (k=1,...,m-4)
\end{equation}
one can show that possible candidates for the role of translation generators
are either $T_{\Lambda}$ or $T'_{\Lambda}$ defined below:
\eqa
T_{0}&=&e_{1}+g_{3}=E_{\epsilon_{l-3}-\epsilon_{l-2}}+E_{\epsilon_{l-3}+\epsilon_{l-2}},
\quad T_{1}=p_{0}-p_{3}=E_{\epsilon_{l-3}+\epsilon_{l-1}}+
E_{\epsilon_{l-3}-\epsilon_{l-1}}\nonumber\\
T'_{0}&=&p_{0}-p_{3}=E_{\epsilon_{l-3}-\epsilon_{l-1}}+
E_{\epsilon_{l-3}+\epsilon_{l-1}}\, ,\,  T'_{1}=q_{1}+q_{2}=-(E_{\epsilon_{l-3}+
\epsilon_{l}}+E_{\epsilon_{l-3}-\epsilon_{l}})
\label{lacosa}
\ena
where $\epsilon_{k},\quad k=1,...,l=\mbox{rank}({\bf so(4,m)})$
is an orthonormal basis of
$\IR^{l}$ and $\epsilon_{i}\pm\epsilon_{j}$ are roots of $so(4,m)$. In the previous
section we found constraints  on the form of the $\IR^{2}$ generators in order for
partial SUSY breaking to occur on a vacuum defined at the origin of $\cQ_{m}$.
 The matrices
$T_{\Lambda}$ fulfill such requirements. On the other hand
one can check that also $T'_{\Lambda}$ are fit to the purpose, even if they do not
 have the form predicted in the previous section.Indeed they cause
the gravitino and gaugino shifts to be proportional
(on the chosen background) to $g_{0}\sigma_{2}+ig_{1}\sigma_{3}$ (and not to
$g_{0}\sigma_{1}+ig_{1}\sigma_{2}$) which, for
$g_{0}=g_{1}$, has also a vanishing eigenvalue.
\par
Taking $T_{\Lambda}$ as generators of $\IR^{2}$, it follows from their matrix form that the largest compact subalgebra of ${\bf so(4,m)}$ suitable for generating $\cG_{compact}$ is ${\bf so(m-1)}$. 
\par
In the previous section it was shown that this choice for the generators $T_{a}$ of $\cG$,
actually
corresponds to an {\it $N=1$  invariant vacuum} defined by the conditions:
\begin{equation}
y^{i}=0 \quad ;\quad b^{a}=0
\end{equation}
 As it was also pointed out before, any point
on $\cQ_{m}$ described by $q=(t,b)$ can define a vacuum on which SUSY
is partially broken, provided that a suitable choice for the isometry generators to be gauged is
done, e.g.:
\eq
T_{\alpha}(t,b)=\IL(t,b)T_{\alpha}\IL(t,b)^{-1}
\label{demografia}
\en
To this extent all the points on the surface spanned by $t^0,t^1$ and containing the origin,
require the same kind of gauging (as it is apparent from
equation (\ref{demografia}), and this is another way of justifying the {\it flatness} of the
scalar potential along these two directions.
\par
The existence of a Killing spinor guarantees that on the corresponding constant scalar
field configuration
$N=2$ SUSY is broken to $N=1$, and the scalar potential vanishes.
Furthermore, as explained in \cite{kilspinold} and already recalled, the existence of
at least one Killing spinor implies
the stability of the background,namely that it is an extremum of
the scalar potential. We have explicitly verified that
the configuration $(y=0;b=0)$ are minima of the scalar potential by plotting
its projection on planes, corresponding to different choices
of pairs $(y,b)$ of coordinates which were let to vary while keeping all the others to zero.
The behaviour of all these curves shows a minimum in the origin, where the
potential vanishes, as expected. Two very typical representatives of the general
behaviour of such  plots are shown in figure 1 and 2.
\begin{figure}
\epsfxsize=5.truecm
\epsfysize=5.truecm
\centerline{\hbox{\epsffile{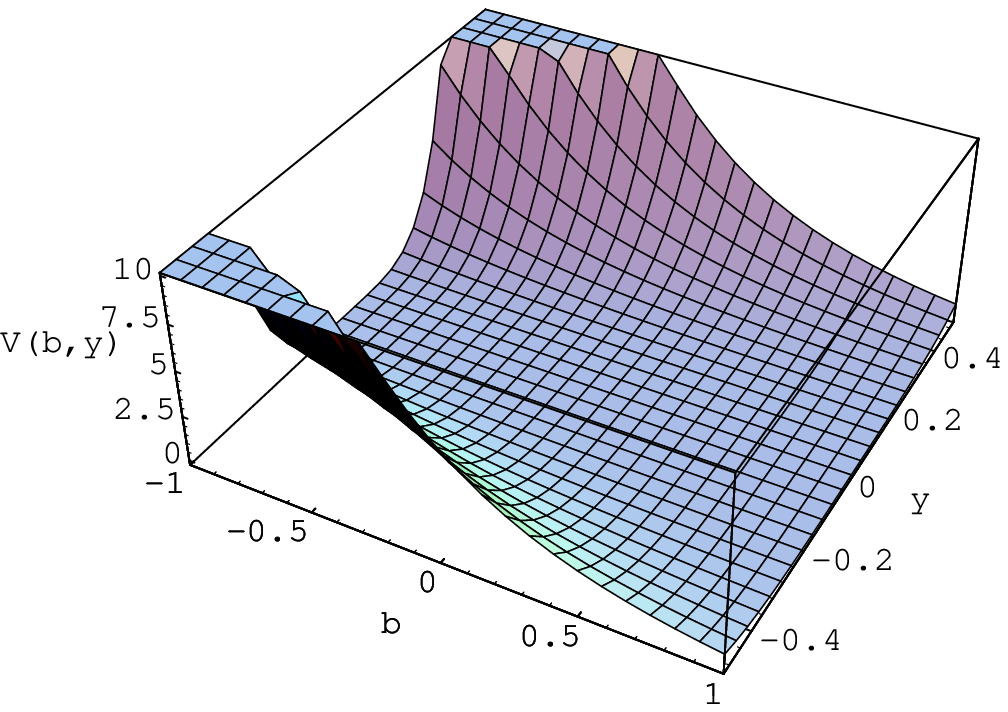}}}
\caption{ Scalar Potential Vs y generic and
b coefficient of $e_0$}
\epsfxsize=5.truecm
\epsfysize=5.truecm
\centerline{\hbox{\epsffile{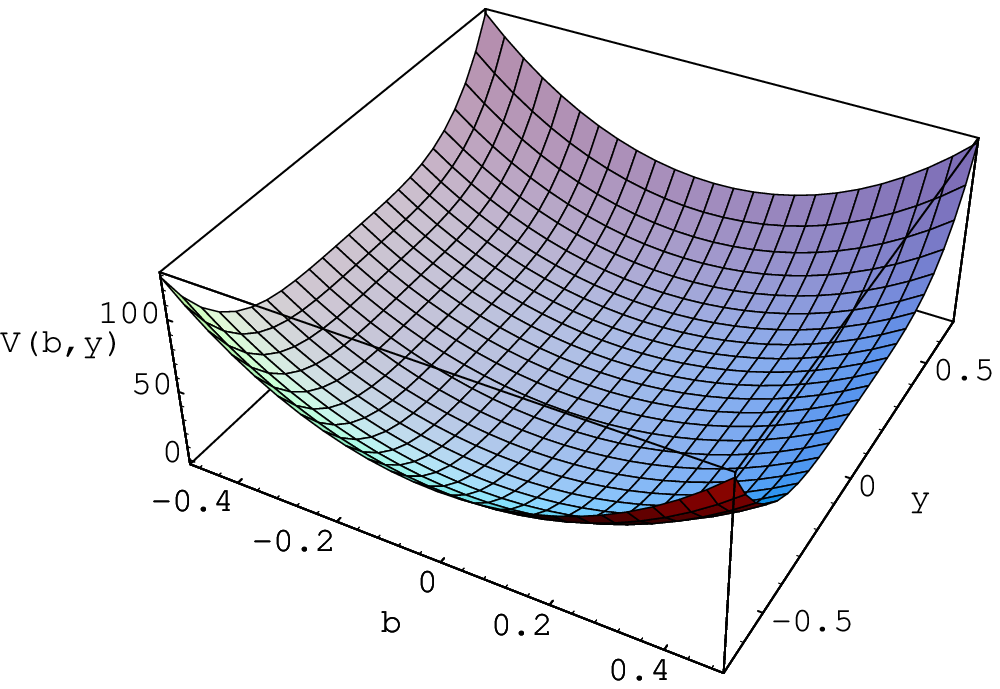}}}
\caption{ Scalar Potential Vs y generic and
b coefficient of $h_1$.}
\end{figure}
\section{Conclusion}
It follows from our analysis that $N=2$ supergravity can be spontaneously broken to $N=1$
supergravity with the survival of unbroken rather arbitrary compact gauge groups.
The fundamental catch in the derivation of the result is the use of the correct
symplectic basis for special geometry and this is the one suggested by the symmetries
of string theory. Here we have described $N=2$ partial breaking
to $N=1$ by means of a classical superHiggs phenomenon in the context of
a {\it microscopic} tree-level supergravity. It is conceivable that the
 further breaking of $N=1$ supersymmetry might proceed
through a totally different mechanism, for instance gaugino condensation.
In order to open new possibilities for phenomenological model building we still have to face the problem of vector like representations of the fermions. We have to look for a further mechanism to break the degeneracy between particles and their N=2 mirrors.
It is also an interesting open question  to
find the relation of our mechanism with the non--perturbative $N=2$
supergravities predicted by string--string
duality and with the conjectured non perturbative breaking caused by extremal black-holes \cite{kalloshlast}.

\end{document}